\documentclass{article}
\usepackage{spconf,amsmath}
\usepackage{graphicx,color,xcolor}
\usepackage{mathtools}
\usepackage{amsthm, amssymb,latexsym}
\usepackage{subfigure}
\usepackage{comment}
\usepackage{hyperref}
\usepackage{booktabs}
\usepackage{url}            
      
\usepackage{microtype}     
\usepackage{lipsum}		
 
\usepackage{amsthm,amssymb,latexsym,amsmath}
\usepackage{amsfonts,mathrsfs}
\usepackage{setspace}
\usepackage{physics}
\usepackage{qcircuit}
\usepackage{braket}
\usepackage{algorithm}
\usepackage{algpseudocode}

\def\argmin{\mathop{\mbox{argmin}}}

\newtheorem{definition}{Definition}[section]
\newtheorem{theorem}{Theorem}[section]

\title{Federated Quantum Machine Learning with Differential Privacy}

\name{Rod Rofougaran$^1$, Shinjae Yoo$^2$, Huan-Hsin Tseng$^2$, Samuel Yen-Chi Chen$^3$\thanks{The views expressed in this article are those of the authors and do not represent the views of Wells Fargo. This article is for informational purposes only. Nothing contained in this article should be construed as investment advice. Wells Fargo makes no express or implied warranties and expressly disclaims all legal, tax, and accounting implications related to this article.}}
\address{$^1$School of Applied and Engineering Physics, Cornell University, Ithaca, NY 14853\\
$^2$Computational Science Initiative, Brookhaven National Laboratory, Upton, NY 11793\\
$^3$Wells Fargo, New York, NY 10017}

\begin{document}
\maketitle

\begin{abstract}
The preservation of privacy is a critical concern in the implementation of artificial intelligence on sensitive training data. There are several techniques to preserve data privacy but quantum computations are inherently more secure due to the no-cloning theorem, resulting in a most desirable computational platform on top of the potential quantum advantages. There have been prior works in protecting data privacy by Quantum Federated Learning (QFL) and Quantum Differential Privacy (QDP) studied independently. However, to the best of our knowledge, no prior work has addressed both QFL and QDP together yet. 
Here, we propose to combine these privacy-preserving methods and implement them on the quantum platform, so that we can achieve comprehensive protection against data leakage (QFL) and model inversion attacks (QDP). This implementation promises more efficient and secure artificial intelligence. In this paper, we present a successful implementation of these privacy-preservation methods by performing the binary classification of the Cats vs Dogs dataset. Using our quantum-classical machine learning model, we obtained a test accuracy of over 0.98, while maintaining epsilon values less than 1.3. We show that federated differentially private training is a viable privacy preservation method for quantum machine learning on Noisy Intermediate-Scale Quantum (NISQ) devices.

\end{abstract}

\section{Introduction}\label{Sec: intro}

Central to the second quantum revolution is the compelling notion that quantum computers possess the potential to achieve exponential speedup over classical counterparts when tackling certain complex problems~\cite{nielsen2010quantum}. An additional intriguing aspect of quantum computations lies in its inherent security advantages. This security originates from the principle of no-cloning~\cite{wootters1982single}, which states that arbitrary unknown quantum states cannot be copied. The implication of this being that an eavesdropper of a quantum computation cannot extract the information of a quantum state without disturbing it. Given these advantages, the integration of quantum machine learning (QML) into the realm of deep learning seems natural. This is particularly relevant as many machine learning tasks demand both strong security measurements and rapid processing of vast datasets. It is worth noting that our approach acknowledges the existing limitations of current quantum hardware, thereby tailoring the proposed quantum computations for execution on NISQ devices~\cite{preskill2018quantum}. To address these constraints. we utilize variational quantum algorithms (VQA)~\cite{bharti2022noisy} that facilitate computation on a limited number of qubits. In the NISQ era, it is well-established that we can leverage noise to our advantage in tackling basic machine learning tasks ~\cite{du2021quantum, yang2023improved}. Our research serves as an illustrative example of this phenomenon. In this study, our goal is to achieve comprehensive security of our learning process. Consequently, we investigate a novel QML approach by joining merits from two distinct privacy-preserving classical techniques: Federated Learning (FL) and Differential Privacy (DP). As a result, we can effectively shield against both model inversion attacks and data leakage, while operating on an inherently secure quantum platform. This paper presents a successful implementation of differentially private federated training on hybrid quantum-classical models. 

\section{Background}\label{Sec: background}
\subsection{Quantum Federated Learning}

Federated Learning (FL)~\cite{mcmahan2017communication} is an acquainted method in processing large amounts of data by parallelization and distribution to multiple computing nodes, which consequently results in the decentralization of training data. This decentralization necessitates the prior partitioning of training data among multiple clients. 
\begin{figure}[htbp]
\begin{center}
\includegraphics[width=\columnwidth]{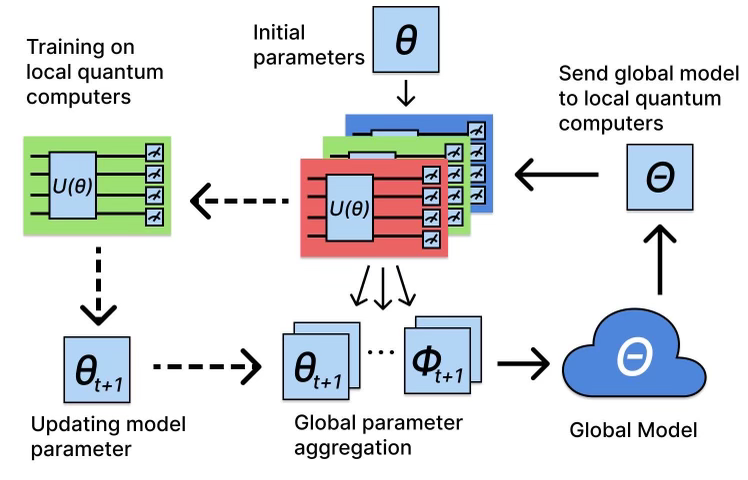}
\vspace*{-7mm}
\caption{{\bfseries The concept of QFL.}}
\label{fig_qfl}
\end{center}
\vskip -0.3in
\end{figure}
The FL cycle begins with a global model $\Theta \in \mathbb{R}^n$ initialized and distributed to $K$ local clients by its identical copies $\Theta_1, \ldots, \Theta_K$, where $\Theta = \Theta_1 \cdots = \Theta_K$ denotes the model parameters to represent the whole model regardless of a \textit{classical} or \textit{quantum} one. Subsequently, client $j \in [K]=\{1, \ldots, K\}$ holding local model $\Theta_j$ engages in local training for a customizable number of epochs to derive a new (private \& local) model $\widetilde{\Theta}_j \neq \Theta_j$. The set of trained client models  $\{\widetilde{\Theta}_j \}_{j=1}^K$ are then aggregated to form a new global model $\widetilde{\Theta}$ and replace initial $\Theta$ to complete one FL cycle. This process occurs iteratively over several rounds. Notably, this training paradigm offers heightened security due to its decentralized nature, which effectively guards against potential data leaks. FL is also advantageous in the context of quantum machine learning because NISQ devices are more suitable for smaller datasets. Without the addition of differential privacy, it has been shown that quantum federated learning can be implemented without any decrease in testing accuracy~\cite{chen2021federated}. The scheme of FL with quantum models is shown in Fig.~\ref{fig_qfl}.

\subsection{Quantum Differential Privacy}
\begin{figure}[htbp]
\begin{center}
\includegraphics[width=0.9\columnwidth]{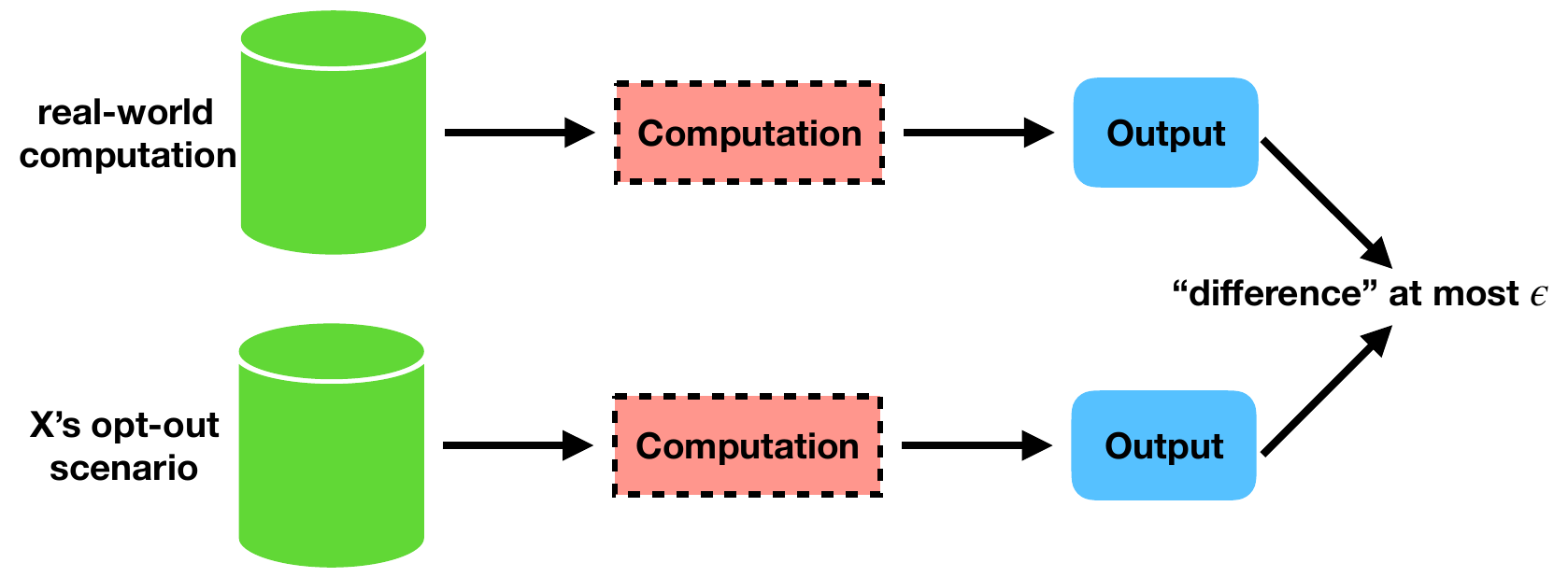}
\caption{{\bfseries The concept of differential privacy.}}
\label{fig_concept_DP}
\end{center}
\vskip -0.3in
\end{figure}

Differential privacy (DP)~\cite{dwork2014algorithmic} aims to fulfill a crucial objective: enabling a data holder to provide assurance to a data subject that, regardless of the insights derived from a conducted study, the confidentiality of the data remains intact. In the context of machine learning, differentially private training ensures that while models can identify general trends in data, they cannot discern individual data points used to train the model. Consequently, differentially private training can effectively protect against model inversion attacks~\cite{abadi2016deep}. It has been shown that quantum differential privacy can safeguard sensitive information while maintaining model accuracy at a satisfactory level~\cite{watkins2023quantum}. Beyond classification tasks, it has also been demonstrated that Quantum Differential Privacy (QDP) algorithms can surpass non-private classical models in sparse regression tasks~\cite{du2022quantum}. The general methodology that upholds this commitment of security to the data subject is illustrated in Fig.~\ref{fig_concept_DP}. Considering two datasets-one with the inclusion of $X$ and one with the exclusion of $X$-it must be ensured that the outputs of these datasets through our models have a bounded difference $\epsilon$. If the difference were not bounded, someone with access to our publicly available model could be able to infer the presence of $X$ in our dataset. 
An $\epsilon$-differential private algorithm is formally defined by Dwork et al ~~\cite{dwork2014algorithmic}. as follows:

\begin{definition}\label{Differential Privacy}
Let $\mathcal{M}$ be a randomized algorithm whose (functional) image is a collection of (probabilistic) \emph{events} $\mathcal{S}$ and the domain is a collection of datasets. If $\mathcal{M}$ is said to be $(\epsilon, \delta)$-differentially private for any dataset $\mathcal{D}_1$, $\mathcal{D}_2$ that differ on a single data point (denoted as $| | \mathcal{D}_1| - | \mathcal{D}_2 | | =1$), we have
     \begin{equation}\label{E: DP}
         Pr[\mathcal{M}(\mathcal{D}_1) \in \mathcal{S}] \leq e^\epsilon \cdot Pr[\mathcal{M}(\mathcal{D}_2) \in \mathcal{S}] + \delta
     \end{equation}

\end{definition}
The quantity $\delta \geq 0$ carries the meaning of \textit{failure probability}~\cite{dwork2014algorithmic}. 
A special case $\delta =0$ is called $\epsilon$-differentially private in which we can observe that $\frac{Pr[\mathcal{M}(\mathcal{D}_1) \in \mathcal{S}]}{ Pr[\mathcal{M}(\mathcal{D}_2) \in \mathcal{S}]} \leq e^\epsilon$. This indicates that when a randomized algorithm $\mathcal{M}$ fails to distinguish two datasets $\mathcal{D}_1$ and $\mathcal{D}_2$, equal probabilities are obtained (or $\epsilon \equiv 0$) to achieve the most private case. Whereas larger $\epsilon \gg 0$ allows two probabilities to be easily distinguished and this results in loss of privacy. Therefore, $\epsilon$ is used to indicate an upper bound on the privacy loss.
A method to make a classifier guarantee differential privacy, Eq.~(\ref{E: DP}), is to add Gaussian noise and gradient clipping within the optimization scheme under the training stage~\cite{abadi2016deep}.
Abadi et al.~\cite{abadi2016deep} also explains how the overall privacy cost is calculated. The study introduces an ''accounting" method referred to as the moment accountant, which accumulates the privacy cost as the training progresses. In their research they provide a proof of the following theorem~\cite{abadi2016deep}:
\begin{theorem}\label{epsilon calculation}
There exists $c_1$ and $c_2$ so that given the number of epochs $T$ and the sampling probability $q = L/N$ where $L$ is the batch size and $N$ is the total number of examples, for any $\epsilon < c_1q^2T$, randomized algorithm $\mathcal{M}$ is $(\epsilon,\delta)$-differentially private for any $\delta > 0$ if we choose the noise level $\sigma$: 
\[
        \sigma \geq \frac{c_2q\sqrt{T \, \log(\frac{1}{\delta})}}{\epsilon}
\]
\end{theorem}
In brief, the value of $\epsilon$ is a function of the following training parameters: the total number of examples, batch size, noise multiplier, number of epochs, and our delta. The primary correlation is, of course, the inverse relationship between $\epsilon$ the noise that we manually input.  

\subsection{Variational Quantum Circuits}
\begin{figure}[htb]
\vskip -0.1in
\begin{center}
\includegraphics[width=0.7\columnwidth]{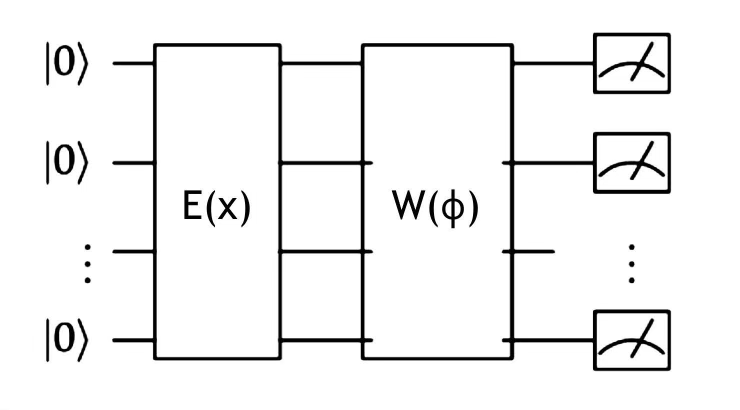}
\vspace*{-3mm}
\caption{{\bfseries A generic structure of a VQC.} A VQC comprises an encoding module denoted as $E(\mathbf{x})$, a trainable component represented as $W(\phi)$, and subsequent measurement operations.}
\label{fig:GenericVQC}
\end{center}
\vskip -0.2in
\end{figure}
Variational quantum circuits (VQC), also known as parameterized quantum circuits (PQC) serve as the quantum counterpart to the classical neural networks. A VQC consists of three major components. The first is the \emph{encoding} part, which can translate a classical vector $\mathbf{x} \in \mathbb{R}^m$ into a quantum state $\ket{\xi}$. We denote the process by an embedding function $\mathbf{x} \mapsto E(\mathbf{x})$ so that $\ket{\xi} = E(\mathbf{x}) \, \ket{0}^{\otimes n}$ (see Fig.~\ref{fig:GenericVQC}). In general, there is no uniform fashion to perform an embedding, where we follow the procedure given in \cite{chen2021federated}.

The \emph{variational} or \emph{learnable} components $W (\phi)$ include multiple single-qubit rotation gates denoted by $W_{ij} (\phi^{(ij)}) = e^{i(\sigma_x \, \alpha_{ij} + \sigma_y \, \beta_{ij} + \sigma_z \, \gamma_{ij} )}$, where $i$ and $j$ represent the index of variational block and qubits, $\phi_{ij} = (\alpha_{ij}, \beta_{ij}, \gamma_{ij})\in \mathbb{R}^3$ are learnable parameters and $\sigma_{k}$ are Pauli matrices. The final measurement operations are to retrieve the information from the circuit for further processing. We utilize the Pauli-$Z$ expectation values in this work. The quantum function can then be defined as $\overrightarrow{f(\mathbf{x} ; \phi)}=\left(\left\langle\hat{Z}_1\right\rangle, \cdots,\left\langle\hat{Z}_N\right\rangle\right)$ , where $\left\langle\hat{Z}_{k}\right\rangle =\left\langle 0\left|E^{\dagger}(\mathbf{x})W^{\dagger}(\phi) \hat{Z_{k}} W(\phi)E(\mathbf{x})\right| 0\right\rangle$. 
By \textit{varying} the parameters $\phi$, the minimization of the objective function can be achieved at $\phi^* = \argmin_{\phi} \mathcal{L}(f(\mathbf{x} ; \phi))$ where $\mathcal{L}$ is the loss function.
The above construction of VQCs offers a multitude of advantages, notably heightened resilience to quantum device noise as evidenced in previous works~\cite{kandala2017hardware,farhi2014quantum,mcclean2016theory}. This attribute proves particularly invaluable in the NISQ era, as highlighted by Preskill~\cite{preskill2018quantum}. In fact, it has been shown by previous works that differential privacy is amplified by the quantum encoding of a classical dataset~\cite{angrisani2022differential}. Additionally, research findings show that VQCs possess a higher level of expressiveness compared to classical neural networks~\cite{sim2019expressibility,lanting2014entanglement,du2018expressive,abbas2021power}. Moreover, they can be effectively trained using smaller datasets, as demonstrated by Caro et al.~\cite{caro2022generalization}. It is noteworthy that VQC applications in QML extend to various domains, including classification~\cite{mitarai2018quantum,qi2023theoretical,chen2021end,chehimi2022quantum,qi2023optimizing,chen2022quantumCNN}, reinforcement learning~\cite{chen2020variational}, natural language processing~\cite{yang2020decentralizing,yang2022bert,di2022dawn,li2023pqlm}, and sequence modeling~\cite{chen2020quantum}.
\section{Methods}\label{Sec:Methods}
\subsection{QFL with DP}
This work integrates the DP and FL in QML through the utilization of VQC. This is achieved by executing DP training on each of our local clients. The global model that is updating and sent to the clients after every iteration, assumes the form of a hybrid quantum-classical machine learning model. The original DP-SGD algorithm ~\cite{abadi2016deep} does not include FL training. We incorporated the original DP-SGD with the FL on VQC models as described in \textbf{Algorithm 1}. We adopted the package PyVacy~\cite{Waites_Pyvacy_2019} to carry out the SGD algorithm and privacy accounting approach for epsilon calculation.
\begin{algorithm}[htb]\label{alg1}
\caption{QFL-DP}
\hspace*{\algorithmicindent} \textbf{Input:}  Examples $\{x_1, \ldots, x_M \}$, loss function $\mathcal{L}(\theta) = \frac{1}{N}\sum_i \mathcal{L}(\theta, x_i)$. \\
\hspace*{\algorithmicindent} \textbf{Parameters:} Clients $K$, selected $J$, local epochs $T$, rounds $R$, learning rate $\eta_t$, noise scale $\sigma$, group size $L$, gradient norm bound $C$. \\
\hspace*{\algorithmicindent} \textbf{Partition:} From $M$ examples, construct $\mathcal{D}_1, \ldots, \mathcal{D}_K$ among $K$ clients randomly, $|\mathcal{D}_i| = N = M/K$ \\
\hspace*{\algorithmicindent} \textbf{Initialize:} Quantum global model $\Theta_0 \in \mathbb{R}^n$ 
\begin{algorithmic}[1]
\For{$r \in [R]$}

\State \textbf{Model distribution:}
\State Make $K$ identical copies of $\Theta_r$ for local set \State $\{\Phi_{r1}, \ldots, \Phi_{rK} \}$ and send $\Phi_{rk}$ to client $k$
\State Take random sample $J$ from $K$ clients
\For{$j \in  [J]$}
\For{$t \in [T]$}
\State \textbf{DP client update:}
\State Perform DP-SGD$(N,\mathcal{L},\eta_t,\sigma,L,C)$ on \State
$\Phi_{rj} \leftarrow \widetilde{\Phi}_{rj} \neq \Phi_{rj}$
\EndFor
\EndFor
\State \textbf{Model aggregation:} \
\State $\Theta_{r+1} = $ averaging the parameters across \State each model in $\{\widetilde{\Phi}_{rj} \}_{j=1}^J$
\EndFor
\end{algorithmic}
\hspace*{\algorithmicindent} \textbf{Output:} $\Theta_R$ and compute the overall privacy cost $(\epsilon, \delta)$ using a privacy accounting method.
\end{algorithm}

\subsection{Hybrid Quantum-Classical Transfer Learning}
\begin{figure}[htb]
\centering
\includegraphics[width=0.9\linewidth]{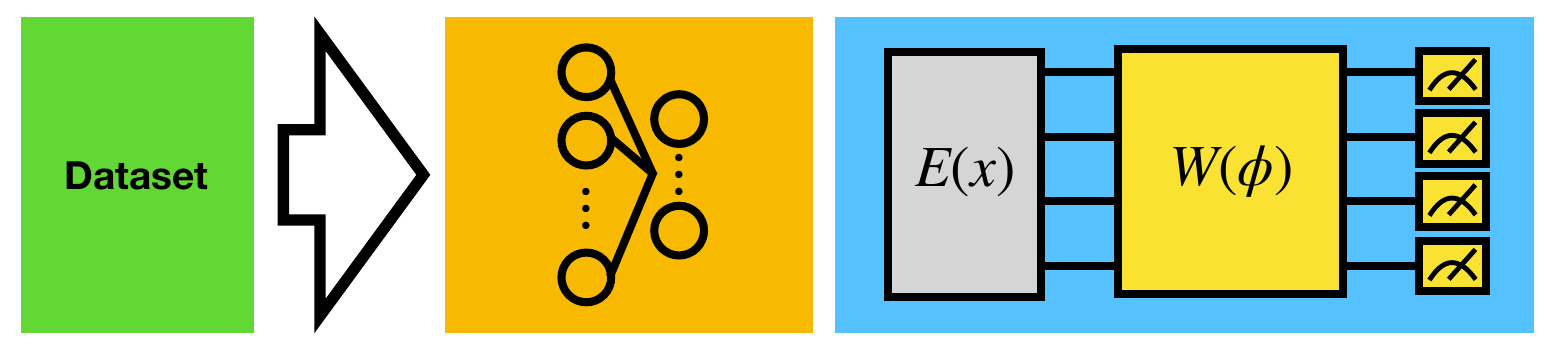}%
\vspace*{-2mm}
\caption{{\bfseries Hybrid Quantum-Classical transfer learning.} }
\label{fig:transfer_learning}
\vskip -0.1in
\end{figure}

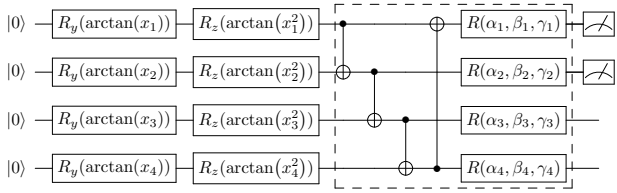
\begin{figure}[htbp]
\centering
\scalebox{0.65}{
\begin{minipage}{10cm}
\Qcircuit @C=1em @R=1em {
\lstick{\ket{0}} & \gate{R_y(\arctan(x_1))} & \gate{R_z(\arctan(x_1^2))} & \ctrl{1}   & \qw       & \qw      & \targ   & \gate{R(\alpha_1, \beta_1, \gamma_1)} & \meter \qw \\
\lstick{\ket{0}} & \gate{R_y(\arctan(x_2))} & \gate{R_z(\arctan(x_2^2))} & \targ      & \ctrl{1}  & \qw      & \qw     & \gate{R(\alpha_2, \beta_2, \gamma_2)} &  \meter \qw \\
\lstick{\ket{0}} & \gate{R_y(\arctan(x_3))} & \gate{R_z(\arctan(x_3^2))} & \qw        & \targ     & \ctrl{1} & \qw     & \gate{R(\alpha_3, \beta_3, \gamma_3)} &  \qw \\
\lstick{\ket{0}} & \gate{R_y(\arctan(x_4))} & \gate{R_z(\arctan(x_4^2))} & \qw        & \qw       & \targ    & \ctrl{-3}& \gate{R(\alpha_4, \beta_4, \gamma_4)} & \qw \gategroup{1}{4}{4}{8}{.7em}{--}\qw 
}
\end{minipage}}
\vspace*{-1mm}
\caption{{\bfseries The VQC used in this work.} }
\label{fig:VQC}
\vskip -0.1in
\end{figure}

The incorporation of a classical component within our model is motivated by the constraints imposed during the NISQ era, where quantum hardware struggles to high fidelity at a large number of qubits or at a substantial circuit depth. Particularly, for computer vision datasets characterized by large data dimensions, such as the Cats vs. Dogs dataset~\cite{asirra-a-captcha-that-exploits-interest-aligned-manual-image-categorization}, the input dimension surpasses the capacity of fully quantum models. Thus, it becomes imperative to integrate a pre-trained classical neural network for input dimensionality reduction prior to feeding it into a VQC~\cite{mari2020transfer} (Fig.~\ref{fig:transfer_learning}).
In this work, we utilize the pre-trained VGG16 model~\cite{simonyan2014very} for dimension reduction. Our model retains VGG16's 16 convolutional layers and integrates a custom classifier that incorporates our VQC. The VQC circuit, as depicted in Fig.~\ref{fig:VQC}, consists of a 4-qubit system and employs a sequence of $R_y$ and $R_z$ gates in the encoding block to transform the input vector $\mathbf{x}$ efficiently.
In the variational block, the qubits are entangled using a series of CNOT gates and followed by the application of general single-qubit unitary gates $R(\alpha,\beta,\gamma)$-controlled by the three learning parameters $\alpha, \beta, \gamma$. The Pauli-$Z$ expectation values of the first two qubits are derived to perform binary classification. The cross-entropy loss function is used in this work.

\section{Experiments \label{sec:experiments}}

\subsection{Experimental Settings}
Our QFL process is initiated by evenly distributing the Cats vs Dogs dataset of 23,000 images among 100 clients. Training occurs in rounds, with randomly selected groups of 5 clients.  At the start of each round, the global model is shared with all clients, but only the chosen 5 perform local SGD training for a set number of epochs. The parameters from these selected clients are aggregated to update the global model for the next round. To validate our framework, we explore different experimental settings, including varying the number of local epochs (1, 2, and 4) and incorporating a non-differentially private model. Each training process is repeated three times to average the outputs and reduce variance. Additionally, we conduct experiments to assess the impact of noise levels during training.

\begin{figure}[htb]
\begin{center}
\includegraphics[width=\columnwidth]{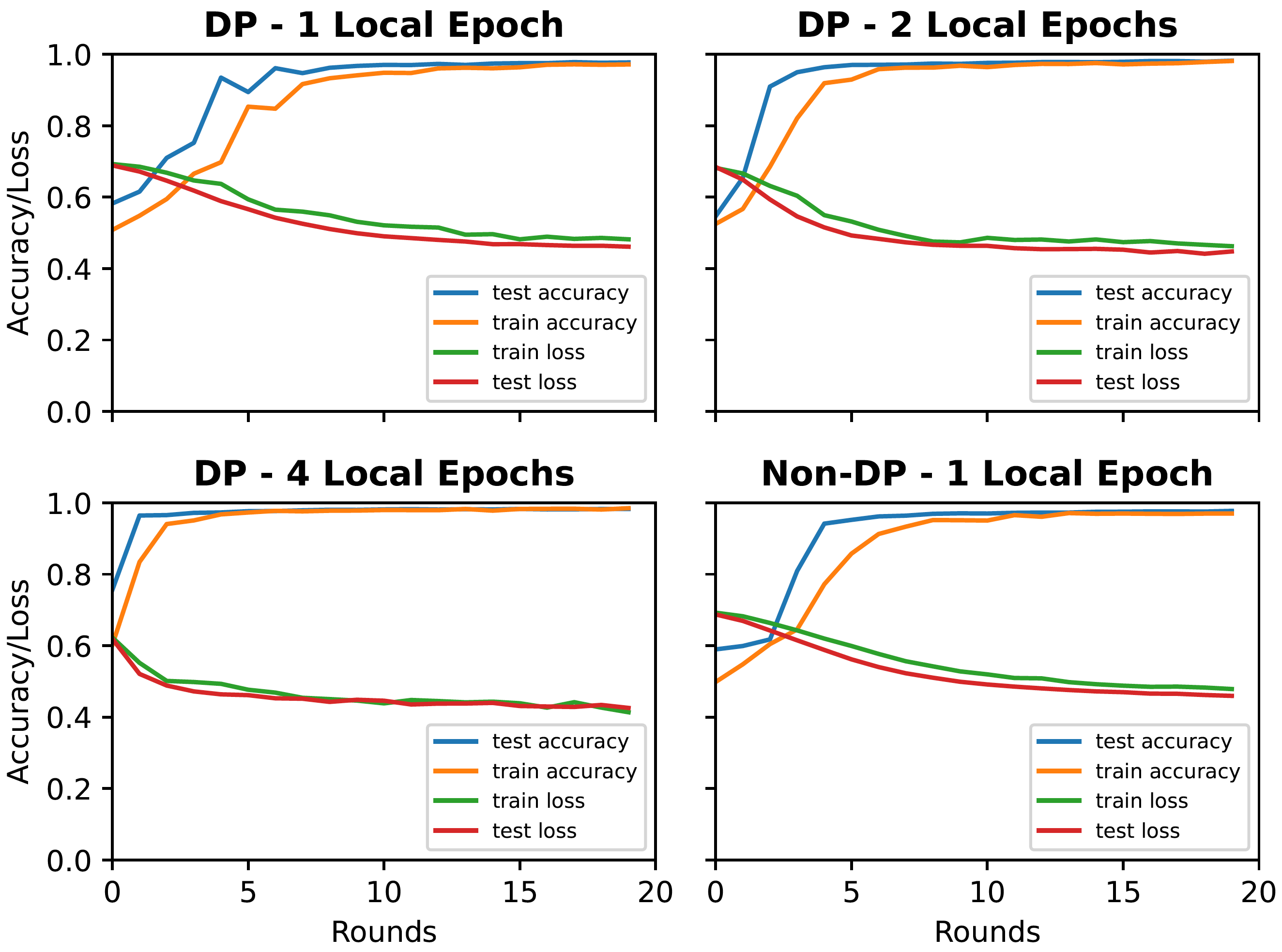}
\vspace*{-4mm}
\caption{All DP plots are $(\epsilon=1.24, \delta=10^{-5})$-DP and acquire test accuracy converging at approximately 0.98.}
\label{fig_res_different_local_epochs}
\end{center}
\vskip -0.1in
\end{figure}

\begin{figure}[htb]
\begin{center}
\includegraphics[width=1.0\columnwidth]{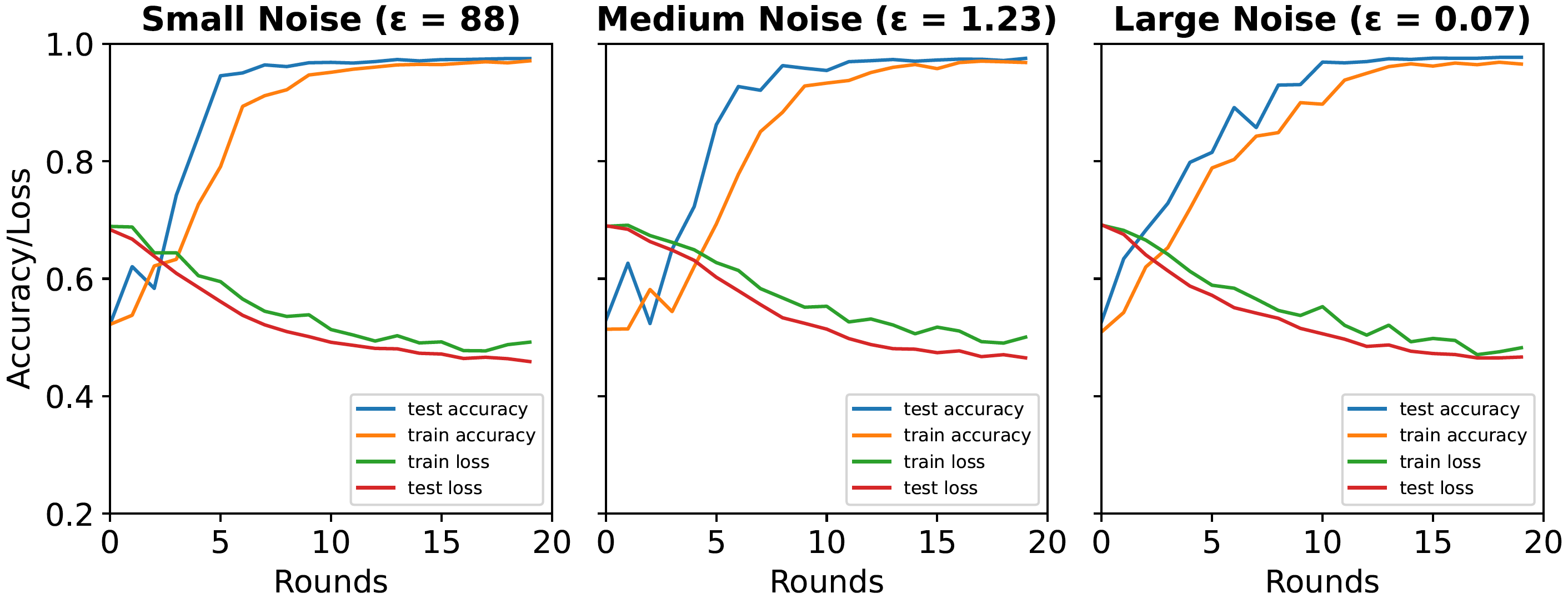}
\vspace*{-6mm}
\caption{[From left to right, $\sigma=0.15, 1, 4$] All plots indicate test accuracy convergence at approximately 0.98.}
\label{fig_res_different_sigma}
\end{center}
\vskip -0.1in
\end{figure}

\subsection{Results}

\textbf{QFL-DP with different local epochs} \quad
We first compare the results of QFL with DP training with various local epochs and the non-DP QFL. The results are shown in Fig.~\ref{fig_res_different_local_epochs}. We observe that all of our models converge to test accuracies of approximately $0.98$ with $\epsilon$'s hovering around $1.24$. It is important to note that the epsilon calculated was the global one, which is a function of total rounds. We also observe that as local epochs increase, a reduction in the number of rounds required to reach convergence, with a decline in variance. Finally, we observe that differentially private training converges slower and with higher variance, which aligns with expectations attributed to the introduction of noise. Additionally, our results are consistent with those of Chen et al.~~\cite{chen2021federated}, which show that the testing accuracy and loss of federated training approximately converge to that of non-federated training. \\
\textbf{QFL-DP with different noise levels} \quad
We further study the correlation between the loss of privacy bound and the accuracy/loss of our models. We study the impact of noise via the increase in $\sigma$ or equivalently the decrease in $\epsilon$. As shown in Fig.~\ref{fig_res_different_sigma}, higher $\epsilon$ results in a slower, higher-variance training process. Generally, increasing the noise enhances privacy but will decrease classification accuracy. However, our results show that the final accuracies of the three cases are not different. Possible reasons are the simplicity of our Cats vs Dogs example and the capabilities of our model architecture.

\section{Conclusion}\label{Sec: conclusion}

Our work demonstrates the effectiveness of differentially private quantum federated learning in mitigating privacy concerns while maintaining competitive performance for NISQ devices. We recognize the need for exploring more complex tasks tailored for quantum algorithms and conducting comparative assessments against classical methods to advance the field of privacy-preserving quantum machine learning.

\clearpage
\begin{spacing}{0.6}
\footnotesize
\bibliographystyle{IEEEbib}
\bibliography{refs,bib/qml_example,bib/vqc,bib/qc_basic,bib/fl,bib/dp,bib/tool}

\begin{thebibliography}{10}

\bibitem{nielsen2010quantum}
Michael~A Nielsen and Isaac~L Chuang,
\newblock ``Quantum computation and quantum information,''
\newblock 2010.

\bibitem{wootters1982single}
William~K Wootters and Wojciech~H Zurek,
\newblock ``A single quantum cannot be cloned,''
\newblock {\em Nature}, vol. 299, no. 5886, pp. 802--803, 1982.

\bibitem{preskill2018quantum}
John Preskill,
\newblock ``Quantum computing in the nisq era and beyond,''
\newblock {\em Quantum}, vol. 2, pp. 79, 2018.

\bibitem{bharti2022noisy}
Kishor Bharti, Alba Cervera-Lierta, Thi~Ha Kyaw, Tobias Haug, Sumner Alperin-Lea, Abhinav Anand, Matthias Degroote, Hermanni Heimonen, Jakob~S Kottmann, Tim Menke, et~al.,
\newblock ``Noisy intermediate-scale quantum algorithms,''
\newblock {\em Reviews of Modern Physics}, vol. 94, no. 1, pp. 015004, 2022.

\bibitem{du2021quantum}
Yuxuan Du, Min-Hsiu Hsieh, Tongliang Liu, Dacheng Tao, and Nana Liu,
\newblock ``Quantum noise protects quantum classifiers against adversaries,''
\newblock {\em Physical Review Research}, vol. 3, no. 2, pp. 023153, 2021.

\bibitem{yang2023improved}
Hang Yang, Xunbo Li, Zhigui Liu, and Witold Pedrycz,
\newblock ``Improved differential privacy noise mechanism in quantum machine learning,''
\newblock {\em IEEE Access}, 2023.

\bibitem{mcmahan2017communication}
Brendan McMahan, Eider Moore, Daniel Ramage, Seth Hampson, and Blaise~Aguera y~Arcas,
\newblock ``Communication-efficient learning of deep networks from decentralized data,''
\newblock in {\em Artificial intelligence and statistics}. PMLR, 2017, pp. 1273--1282.

\bibitem{chen2021federated}
Samuel Yen-Chi Chen and Shinjae Yoo,
\newblock ``Federated quantum machine learning,''
\newblock {\em Entropy}, vol. 23, no. 4, pp. 460, 2021.

\bibitem{dwork2014algorithmic}
Cynthia Dwork, Aaron Roth, et~al.,
\newblock ``The algorithmic foundations of differential privacy,''
\newblock {\em Foundations and Trends{\textregistered} in Theoretical Computer Science}, vol. 9, no. 3--4, pp. 211--407, 2014.

\bibitem{abadi2016deep}
Martin Abadi, Andy Chu, Ian Goodfellow, H~Brendan McMahan, Ilya Mironov, Kunal Talwar, and Li~Zhang,
\newblock ``Deep learning with differential privacy,''
\newblock in {\em Proceedings of the 2016 ACM SIGSAC conference on computer and communications security}, 2016, pp. 308--318.

\bibitem{watkins2023quantum}
William~M Watkins, Samuel Yen-Chi Chen, and Shinjae Yoo,
\newblock ``Quantum machine learning with differential privacy,''
\newblock {\em Scientific Reports}, vol. 13, no. 1, pp. 2453, 2023.

\bibitem{du2022quantum}
Yuxuan Du, Min-Hsiu Hsieh, Tongliang Liu, Shan You, and Dacheng Tao,
\newblock ``Quantum differentially private sparse regression learning,''
\newblock {\em IEEE Transactions on Information Theory}, vol. 68, no. 8, pp. 5217--5233, 2022.

\bibitem{kandala2017hardware}
Abhinav Kandala, Antonio Mezzacapo, Kristan Temme, Maika Takita, Markus Brink, Jerry~M Chow, and Jay~M Gambetta,
\newblock ``Hardware-efficient variational quantum eigensolver for small molecules and quantum magnets,''
\newblock {\em Nature}, vol. 549, no. 7671, pp. 242--246, 2017.

\bibitem{farhi2014quantum}
Edward Farhi, Jeffrey Goldstone, and Sam Gutmann,
\newblock ``A quantum approximate optimization algorithm,''
\newblock {\em arXiv preprint arXiv:1411.4028}, 2014.

\bibitem{mcclean2016theory}
Jarrod~R McClean, Jonathan Romero, Ryan Babbush, and Al{\'a}n Aspuru-Guzik,
\newblock ``The theory of variational hybrid quantum-classical algorithms,''
\newblock {\em New Journal of Physics}, vol. 18, no. 2, pp. 023023, 2016.

\bibitem{angrisani2022differential}
Armando Angrisani, Mina Doosti, and Elham Kashefi,
\newblock ``Differential privacy amplification in quantum and quantum-inspired algorithms,''
\newblock {\em arXiv preprint arXiv:2203.03604}, 2022.

\bibitem{sim2019expressibility}
Sukin Sim, Peter~D Johnson, and Al{\'a}n Aspuru-Guzik,
\newblock ``Expressibility and entangling capability of parameterized quantum circuits for hybrid quantum-classical algorithms,''
\newblock {\em Advanced Quantum Technologies}, vol. 2, no. 12, pp. 1900070, 2019.

\bibitem{lanting2014entanglement}
Trevor Lanting, Anthony~J Przybysz, A~Yu Smirnov, Federico~M Spedalieri, Mohammad~H Amin, Andrew~J Berkley, Richard Harris, Fabio Altomare, Sergio Boixo, Paul Bunyk, et~al.,
\newblock ``Entanglement in a quantum annealing processor,''
\newblock {\em Physical Review X}, vol. 4, no. 2, pp. 021041, 2014.

\bibitem{du2018expressive}
Yuxuan Du, Min-Hsiu Hsieh, Tongliang Liu, and Dacheng Tao,
\newblock ``The expressive power of parameterized quantum circuits,''
\newblock {\em arXiv preprint arXiv:1810.11922}, 2018.

\bibitem{abbas2021power}
Amira Abbas, David Sutter, Christa Zoufal, Aur{\'e}lien Lucchi, Alessio Figalli, and Stefan Woerner,
\newblock ``The power of quantum neural networks,''
\newblock {\em Nature Computational Science}, vol. 1, no. 6, pp. 403--409, 2021.

\bibitem{caro2022generalization}
Matthias~C Caro, Hsin-Yuan Huang, Marco Cerezo, Kunal Sharma, Andrew Sornborger, Lukasz Cincio, and Patrick~J Coles,
\newblock ``Generalization in quantum machine learning from few training data,''
\newblock {\em Nature communications}, vol. 13, no. 1, pp. 1--11, 2022.

\bibitem{mitarai2018quantum}
Kosuke Mitarai, Makoto Negoro, Masahiro Kitagawa, and Keisuke Fujii,
\newblock ``Quantum circuit learning,''
\newblock {\em Physical Review A}, vol. 98, no. 3, pp. 032309, 2018.

\bibitem{qi2023theoretical}
Jun Qi, Chao-Han~Huck Yang, Pin-Yu Chen, and Min-Hsiu Hsieh,
\newblock ``Theoretical error performance analysis for variational quantum circuit based functional regression,''
\newblock {\em npj Quantum Information}, vol. 9, no. 1, pp. 4, 2023.

\bibitem{chen2021end}
Samuel Yen-Chi Chen, Chih-Min Huang, Chia-Wei Hsing, and Ying-Jer Kao,
\newblock ``An end-to-end trainable hybrid classical-quantum classifier,''
\newblock {\em Machine Learning: Science and Technology}, vol. 2, no. 4, pp. 045021, 2021.

\bibitem{chehimi2022quantum}
Mahdi Chehimi and Walid Saad,
\newblock ``Quantum federated learning with quantum data,''
\newblock in {\em ICASSP 2022-2022 IEEE International Conference on Acoustics, Speech and Signal Processing (ICASSP)}. IEEE, 2022, pp. 8617--8621.

\bibitem{qi2023optimizing}
Jun Qi, Xiao-Lei Zhang, and Javier Tejedor,
\newblock ``Optimizing quantum federated learning based on federated quantum natural gradient descent,''
\newblock in {\em ICASSP 2023-2023 IEEE International Conference on Acoustics, Speech and Signal Processing (ICASSP)}. IEEE, 2023, pp. 1--5.

\bibitem{chen2022quantumCNN}
Samuel Yen-Chi Chen, Tzu-Chieh Wei, Chao Zhang, Haiwang Yu, and Shinjae Yoo,
\newblock ``Quantum convolutional neural networks for high energy physics data analysis,''
\newblock {\em Physical Review Research}, vol. 4, no. 1, pp. 013231, 2022.

\bibitem{chen2020variational}
Samuel Yen-Chi Chen, Chao-Han~Huck Yang, Jun Qi, Pin-Yu Chen, Xiaoli Ma, and Hsi-Sheng Goan,
\newblock ``Variational quantum circuits for deep reinforcement learning,''
\newblock {\em IEEE Access}, vol. 8, pp. 141007--141024, 2020.

\bibitem{yang2020decentralizing}
Chao-Han~Huck Yang, Jun Qi, Samuel Yen-Chi Chen, Pin-Yu Chen, Sabato~Marco Siniscalchi, Xiaoli Ma, and Chin-Hui Lee,
\newblock ``Decentralizing feature extraction with quantum convolutional neural network for automatic speech recognition,''
\newblock in {\em ICASSP 2021-2021 IEEE International Conference on Acoustics, Speech and Signal Processing (ICASSP)}. IEEE, 2021, pp. 6523--6527.

\bibitem{yang2022bert}
Chao-Han~Huck Yang, Jun Qi, Samuel Yen-Chi Chen, Yu~Tsao, and Pin-Yu Chen,
\newblock ``When bert meets quantum temporal convolution learning for text classification in heterogeneous computing,''
\newblock in {\em ICASSP 2022-2022 IEEE International Conference on Acoustics, Speech and Signal Processing (ICASSP)}. IEEE, 2022, pp. 8602--8606.

\bibitem{di2022dawn}
Riccardo Di~Sipio, Jia-Hong Huang, Samuel Yen-Chi Chen, Stefano Mangini, and Marcel Worring,
\newblock ``The dawn of quantum natural language processing,''
\newblock in {\em ICASSP 2022-2022 IEEE International Conference on Acoustics, Speech and Signal Processing (ICASSP)}. IEEE, 2022, pp. 8612--8616.

\bibitem{li2023pqlm}
Shuyue~Stella Li, Xiangyu Zhang, Shu Zhou, Hongchao Shu, Ruixing Liang, Hexin Liu, and Leibny~Paola Garcia,
\newblock ``Pqlm-multilingual decentralized portable quantum language model,''
\newblock in {\em ICASSP 2023-2023 IEEE International Conference on Acoustics, Speech and Signal Processing (ICASSP)}. IEEE, 2023, pp. 1--5.

\bibitem{chen2020quantum}
Samuel Yen-Chi Chen, Shinjae Yoo, and Yao-Lung~L Fang,
\newblock ``Quantum long short-term memory,''
\newblock in {\em ICASSP 2022-2022 IEEE International Conference on Acoustics, Speech and Signal Processing (ICASSP)}. IEEE, 2022, pp. 8622--8626.

\bibitem{Waites_Pyvacy_2019}
C.~Waites,
\newblock ``Pyvacy: Privacy algorithms for pytorch,''
\newblock https://github.com/ChrisWaites/pyvacy, 2019.

\bibitem{asirra-a-captcha-that-exploits-interest-aligned-manual-image-categorization}
Jeremy Elson, John~(JD) Douceur, Jon Howell, and Jared Saul,
\newblock ``Asirra: A captcha that exploits interest-aligned manual image categorization,''
\newblock in {\em Proceedings of 14th ACM Conference on Computer and Communications Security (CCS)}. October 2007, Association for Computing Machinery, Inc.

\bibitem{mari2020transfer}
Andrea Mari, Thomas~R Bromley, Josh Izaac, Maria Schuld, and Nathan Killoran,
\newblock ``Transfer learning in hybrid classical-quantum neural networks,''
\newblock {\em Quantum}, vol. 4, pp. 340, 2020.

\bibitem{simonyan2014very}
Karen Simonyan and Andrew Zisserman,
\newblock ``Very deep convolutional networks for large-scale image recognition,''
\newblock {\em arXiv preprint arXiv:1409.1556}, 2014.

\end{thebibliography}
\end{spacing}

\end{document}